\documentclass[11pt]{article}

\usepackage{amsxtra,amssymb,amsthm,amsmath,latexsym}
\usepackage{graphicx}
\usepackage{epsfig}
\usepackage{afterpage}

\newtheorem{lem}[]{Lemma}

 \newcommand{\lemref}[1]{Lemma~\ref{#1}}

\newcommand{\R}{{\mathbb R}}

\newcommand{\nb}{\nabla}

\newcommand{\dl}{{\delta}}
\newcommand{\bee}{\begin{equation*}}
\newcommand{\eee}{\end{equation*}}
\newcommand{\be}{\begin{equation}}
\newcommand{\ee}{\end{equation}}

\title{Electromagnetic wave scattering by a thin layer in which many small 
particles are embedded}
\author{A.G. Ramm \\
\small Department of Mathematics\\[-0.8ex]
\small Kansas State University, Manhattan, KS 66506-2602, USA\\
\small \texttt{ramm@math.ksu.edu}}

\date{}
\begin{document}

Progress in Electromagnetic Research Letters, 19, (2010), 147-154.
\maketitle

\begin{abstract}
Scattering of electromagnetic (EM) waves by many small particles (bodies),
embedded in a thin layer, is studied. Physical properties of
the particles are described by their boundary impedances.
The thin layer of depth of the order $O(a)$  with many embedded in it 
small particles of characteristic size $a$,  is 
described by a boundary condition on the surface of the layer. The
limiting interface boundary condition is  obtained
for the effective EM field in the limiting medium, 
in the limit $a\to 0$, where 
the number $M(a)$ of the particles tends to infinity at
a suitable rate. 
\end{abstract}
 



\section{Introduction}
It is known (see, e.g., \cite{MK}, \cite{S})  that the light propagation 
through
diffraction gratings may exibit strong resonances at certain frequencies.
This is useful in applications. In this paper we study electromagnetic 
(EM) wave scattering by many small impedance particles $D_m$,
$1\leq m \leq M$, $M=M(a)$,  embedded 
in a thin layer of the depth $h(a)\sim a$, where $a$ is the characteristic 
dimension of a small particle. The shape of the particles may be fairly 
arbitrary, not necessarily spherical. 

We assume that
\be\label{e1}
\lim_{a\to 0} a/d(a) =0, \qquad lim_{a\to 0}d(a) =0, \qquad ka\ll 1,
\ee
where $k$ is the wavenumber, and $d(a)$ is the distance between 
neighboring particles.   

The thin layer is located on a smooth surface $S$. 
The permettivity $\epsilon_0$, conductivity $\sigma_0\ge 0$, 
$\epsilon'=\epsilon_0+\frac {i\sigma_0}{\omega}$ and permeability 
$\mu_0$ of the space  are 
assumed constants, $k^2=\omega^2\epsilon'\mu_0$, $\omega$ is the 
frequency.

For example,  one may 
assume that $S$ is the plane $x_3=0$, but our arguments are valid 
for an arbitrary smooth $S$. 
The  $M$ particles on $S$ are distributed according to the
following law: in any open subset $\Delta$ of $S$ there are 
$\mathcal{N}(\Delta)$ particles, where
\be\label{e2}
\mathcal{N}(\Delta)=\frac{1}{a^{2-\kappa}}\int_{\Delta}N(s)ds[1+o(1)],\quad
a\to 0, 
\ee
$N(s)\geq 0$ is a continuous function, vanishing
outside of a finite domain $\Omega\subset S$ in which
small particles (bodies)  $D_m$ are distributed,
$\kappa\in(0,1)$ is a number, and
the boundary impedances of the small particles are defined by the
formula
\be\label{eIm}
\zeta_m=\frac{h(s_m)}{a^\kappa},\quad s_m\in D_m,
\ee
where $s_m\in S$ is a point inside $m-$th particle $D_m$, Re $h(s)\geq 0$, 
and
$h(s)$ is a continuous function vanishing
outside $\Omega.$  

{\it We can choose $\kappa$ and $h(s)$ as we wish.}  

Denote by $[E,H]=E\times H$ the cross
product of two vectors, and by $(E,H)=E\cdot H$ the dot product of two
vectors.

The 
impedance boundary condition on the surface $S_m$
of the $m-$th particle $D_m$ is $E^t=\zeta_m [H^t,N]$, where $E^t$ ($H^t$)
is the tangential component of $E$ ($H$) on $S_m$, and $N$ is the unit
normal to $S_m$, pointing out of $D_m$. We define  $E^t=[N,[E,N]]=E-N(E,N)$.  
This corresponds to the geometrical meaning of the tangential component of 
$E$, and differs from the definition  $E^t=[N,E]$ that is used sometimes.

In this paper we use the methodology, developed in \cite{R581} and 
some results from \cite{R581}. 

\section{EM wave scattering by many small particles}

Electromagnetic (EM) wave scattering problem
consists of finding vectors $E$ and $H$ satisfying the Maxwell
equations: 
\be\label{e3} \nb \times E=i\omega\mu_0 H,\quad \nb\times
H=-i\omega \epsilon_0 E\quad \text{in } D:=\R^3\setminus
\cup_{m=1}^M D_m, \ee 
the impedance boundary conditions:
\be\label{e4} [N,[E,N]]=\zeta_m[H,N]\text{ on
} S_m,\ 1\leq m\leq M, \ee
and the radiation conditions:
\be\label{e5} E=E_0+v_E,\quad H=H_0+v_H,
 \ee
where $v_E$ and $v_H$ satify the radiation condition, $\zeta_m$ is the 
impedance, defined in \eqref{eIm}, $N$ is the 
unit normal to $S_m$
pointing out of $D_m$, $E_0, H_0$ are the incident fields satisfying
equations \eqref{e1} in all of $\R^3$. One often assumes that
the incident wave is a plane wave, i.e., 
$E_0=\mathcal{E}e^{ik\alpha\cdot x}$, $\mathcal{E}$
is a constant vector, $\alpha\in S^2$ is a unit vector, $S^2$ is the
unit sphere in $\R^3$, $\alpha\cdot \mathcal{E}=0$, 
$v_E $ and $v_H$ satisfy the
radiation condition: $r(\frac{\partial v}{\partial
r}-ikv)=o(1)$ as $r:=|x|\to \infty$, $k=\omega\sqrt{\epsilon_0 \mu_0}$.

By impedance $\zeta_m$ we assume in this paper a constant,
Re $\zeta_m\geq 0$, 
or a matrix function $2\times 2$ acting on the
tangential to $S_m$ vector fields, such that 
\be\label{e6}
\text{Re}(\zeta_mE^t,E^t)\geq 0\quad \forall E^t\in T_m, \ee where
$T_m$ is the set of all tangential to $S_m$ continuous vector fields
such that Div$E^t=0$, where Div is the surface divergence, and $E^t$
is the tangential component of $E$. 
Smallness
of $D_m$ means that $ka\ll 1$, where $a=0.5\max_{1\leq m\leq M}
\text{diam} D_m$. 

\begin{lem}\label{lem1}
Problem \eqref{e3}-\eqref{e6} has at most one solution.
\end{lem}
\lemref{lem1} is proved in \cite{R581}.\\
Let us note that problem \eqref{e3}-\eqref{e6} is equivalent to the
problems \eqref{e7}, \eqref{e8}, \eqref{e5}, \eqref{e6}, where
\be\label{e7} \nb\times \nb\times E=k^2E\text{ in } D,\qquad
H=\frac{\nb\times E}{i\omega \mu_0}, \ee 
\be\label{e8}
[N,[E,N]]=\frac{\zeta_m}{i\omega \mu_0}[\nb\times E,N]\text{ on }
S_m,\ 1\leq m\leq M. \ee 
This is the impedance boundary condition (see, e.g., \cite{LL}, p. 301.)
The expression $[N,[E,N]]=E-(E,N)N$ is the tangential component of
the field $E$ on the surface $S_m$, $N$ is the unit normal to $S_m$
pointing out of $D_m$.

Thus, we have reduced our problem to
finding one vector $E(x)$. If $E(x)$ is found, then
$H=\frac{\nb\times E}{i\omega\mu_0}.$ 

Let us look for $E$ of the
form 
\be\label{e9} E=E_0+\sum_{m=1}^M\nb \times
\int_{S_m}g(x,t)\sigma_m(t)dt,\quad
g(x,y)=\frac{e^{ik|x-y|}}{4\pi|x-y|}, \ee 
where $t\in S_m$,  $dt$
is an element of the area of $S_m$, and $\sigma_m(t)\in T_m$. This $E$
for any continuous $\sigma_m(t)$ solves equation \eqref{e7} in $D$
because $E_0$ solves \eqref{e7} and 
\be\label{e10}\begin{split}
\nb\times\nb\times\nb\times \int_{S_m}g(x,t)\sigma_m(t)dt&=\nb \nb\cdot
\nb\times\int_{S_m}g(x,t)\sigma_m(t)dt\\
&-\nb^2\nb\times
\int_{S_m}g(x,t)\sigma_m(t)dt\\
&=k^2\nb\times \int_{S_m}g(x,t)\sigma_m(t)dt,\quad x\in D.
\end{split}\ee
Here we have used the known identity $div curl E=0,$ valid
for any smooth vector field $E$, and the known formula 
\be\label{eG}
-\nb^2 g(x,y)=k^2g(x,y)+\dl(x-y). 
\ee
The integral
$\int_{S_m}g(x,t)\sigma_m(t)dt$ satisfies the radiation condition.
Thus, formula \eqref{e9} gives $E(x)$ that solves problem \eqref{e7}, 
\eqref{e8},
and satisfies the radiation condition, if $\sigma_m(t)$ are chosen so that 
boundary
conditions \eqref{e8} are satisfied. 

Define the effective field
$E_e(x)=E_e^m(x)=E_e^{(m)}(x,a),$ acting on the $m-$th body $D_m$:
\be\label{e11} E_e(x):=E(x)-\nabla\times
\int_{S_m}g(x,t)\sigma_m(t)dt:=E_e^{(m)}(x), \ee where we assume
that $x$ is in a neigborhood of $S_m$. However,  $E_e(x)$ is defined 
for all $x\in \R^3$. 

Away from $S$, the field $E_e(x,a)$ tends to a limit
$E(x)=E_e(x)$ as $a\to 0$, and $E_e(x)$ is a twice continuously
differentiable function away from $S$, see \cite{R581}.
 To derive an integral equation for
$\sigma_m=\sigma_m(t)$, substitute
$$E=E_e+\nb\times\int_{S_m}g(x,t)\sigma_m(t)dt$$ into boundary condition  
\eqref{e8}, use the known formula (see, e.g., \cite{M}) 
\be\label{e12} [N,\nb\times
\int_{S_m}g(x,t)\sigma_m(t)dt]_{\mp}=\int_{S_m}
[N_s,[\nb_sg(x,t),\sigma_m(t)]]dt\pm \frac{\sigma_m(t)}{2}, \ee 
where the -(+) signs
denote the limiting values of the left-hand side of \eqref{e12} as $x\to 
s$ from $D$ $(D_m)$, and get 
\be\label{e13} \sigma_m(t)=A_m\sigma_m+f_m,\quad
1\leq m\leq M. \ee 
Here $A_m$ is a linear Fredholm-type integral
operator, and $f_m$ is a continuously differentiable function. 

Let us specify $A_m$ and $f_m$. One has 
\be\label{e14}
f_m=2[N_s,f_e(s)],\quad
f_e(s):=[N_s,[E_e(s),N_s]]-\frac{\zeta_m}{i\omega \mu_0}[\nb\times
E_e,N_s]. \ee 
Condition \eqref{e8} and formula \eqref{e12} yield
\be\label{e15}\begin{split}
&f_e(s)+\frac{1}{2}[\sigma_m(s),N_s]+
[\int_{S_m}[N_s,[\nb_sg(s,t),\sigma_m(t)]]dt,N_s]\\
&-\frac{\zeta_m}{i\omega
\mu_0}[\nb\times\nb\times\int_{S_m}g(x,t)\sigma_m(t)dt,N_s]|_{x\to
s}=0\end{split}\ee

Using the known formula $\nb\times\nb=grad div  -\nb^2$, the relation
\be\label{e16}\begin{split}
\nb_x\nb_x \cdot\int_{S_m}g(x,t)\sigma_m(t)dt&=\nb_x\int_{S_m}(-\nb_t
g(x,t),\sigma_m(t))dt\\
&=\nb_x\int_{S_m}g(x,t)\text{Div}\sigma_m(t)dt=0, \end{split}\ee
where Div is the surface divergence, and a consequence of formula 
\eqref{eG} 
\be\label{e17}
-\nb_x^2\int_{S_m}g(x,t)\sigma_m(t)dt=k^2\int_{S_m}g(x,t)\sigma_m(t)dt,\quad
x\notin S, \ee 
one gets from \eqref{e15} the following equation
\be\label{e18} [N_s,\sigma_m(s)]+2f_e(s)+2B\sigma_m=0. \ee 
Here
\be\label{e19} B\sigma_m:=[\int_{S_m}[N_s,[\nb_s
g(s,t),\sigma_m(t)]]dt,N_s]+\zeta_mi\omega
\epsilon_0[\int_{S_m}g(s,t)\sigma_m(t)dt,N_s]. \ee 
Take cross
product of $N_s$ with the left-hand side of \eqref{e18} and use the
formulas $N_s\cdot \sigma_m(s)=0$, $f_m:=f_m(s):=2[N_s, f_e(s)]$, and 
\be\label{e20}
[N_s,[N_s,\sigma_m(s)]]=-\sigma_m(s), \ee to get from \eqref{e18}
equation \eqref{e13}: 
\be\label{e21}
\sigma_m(s)=2[N_s,f_e(s)]+2[N_s,B\sigma_m]:=A_m\sigma_m+f_m, \ee
where $A_m\sigma_m=2[N_s,B\sigma_m]$. The operator $A_m$ is linear and 
compact
in the space $C(S_m)$, so that equation \eqref{e21} is of Fredholm
type. Therefore, equation \eqref{e21} is solvable for any $f_m\in
T_m$ if the homogeneous version of \eqref{e21} has only the trivial
solution $\sigma_m=0$. In this case the solution $\sigma_m$ 
to equation \eqref{e21} is of the 
order of the right-hand side $f_m$, that is, $O(a^{-\kappa})$ as $a\to 
0$, see formula \eqref{e14}. Moreover, it follows from equation 
\eqref{e21} that the main term of the asymptotics of $\sigma_m$ 
as $a\to 0$ does not depend on $s\in S_m$.

\begin{lem}\label{lem2}
Assume that $\sigma_m\in T_m, $ $\sigma_m\in C(S_m)$, and
$\sigma_m(s)=A_m\sigma_m$. Then $\sigma_m=0$.
\end{lem}
\lemref{lem2} is proved in \cite{R581}.

Let us write \eqref{e9} as 
\be\label{e22}
E(x)=E_0(x)+\sum_{m=1}^M[\nb_x g(x,x_m),Q_m]+\sum_{m=1}^M
\nb\times\int_{S_m}(g(x,t)-g(x,x_m))\sigma_m(t)dt, \ee where
\be\label{e23} Q_m:=\int_{S_m}\sigma_m(t)dt. 
\ee
Since $\sigma_m=O(a^{-\kappa})$, one has $Q_m=O(a^{2-\kappa})$. 
We want to prove
that the second sum in \eqref{e22} is negligible compared with the 
first sum. 
One has
\be\label{e24} j_1:=|[\nb_x g(x,x_m),Q_m]|\leq
O\left(\max\left\{\frac{1}{d^2},\frac{k}{d}\right\}\right)O(a^{2-\kappa}),
\ee 
\be\label{e25}
j_2:=|\nb\times\int_{S_m}(g(x,t)-g(x,x_m))\sigma_m(t)dt|\leq a
O\left(\max\left\{\frac{1}{d^3},\frac{k^2}{d}\right\}\right)O(a^{2-\kappa}),
\ee 
and 
\be\label{e26} \left| \frac{j_2}{j_1}\right|=O\left( \max
\left\{ \frac{a}{d},ka\right\}\right)\to 0,\qquad \frac a d=o(1),  \qquad 
a\to 0.  \ee
Thus, one may neglect the second sum in \eqref{e24}, and write
\be\label{e27} E(x)=E_0(x)+\sum_{m=1}^M[\nb_xg(x,x_m),Q_m] \ee with
an error that tends to zero as $a\to 0$. Let us estimate $Q_m$
asymptotically, as $a\to 0$. Integrate equation \eqref{e21} over
$S_m$ to get 
\be\label{e28}
Q_m=2\int_{S_m}[N_s,f_e(s)]ds+2\int_{S_m}[N_s,B\sigma_m]ds .\ee It
follows from \eqref{e14} that 
\be\label{e29}
[N_s,f_e]=[N_s,E_e]-\frac{\zeta_m}{i\omega \mu_0}[N_s,[\nb \times
E_e,N_s]]. \ee If $E_e$ tends to a finite limit as $a\to 0$, then
formula \eqref{e29} implies 
\be\label{e30}
[N_s,f_e]=O(\zeta_m)=O\left(\frac{1}{a^\kappa}\right),\quad a\to 0.
\ee By \lemref{lem2} the operator $(I-A_m)^{-1}$ is bounded, so
$\sigma_m=O\left(\frac{1}{a^\kappa}\right)$, and 
\be\label{e31}
Q_m=O\left(a^{2-\kappa}\right),\quad a\to 0, \ee because integration
over $S_m$ adds factor $O(a^2)$. As $a\to 0$,
the sum \eqref{e27} converges to the integral (see \cite{R586}, Lemma 1) 
\be\label{e32}
E=E_0+\nb\times\int_{S}g(x,s)N(s)Q(s)ds, \ee 
where $N(s)$ is the function from \eqref{e2}, and $Q(s)$ is the function 
such that 
\be\label{e33} Q_m=Q(x_m)a^{2-\kappa}. \ee
The function $Q(y)$ can be expressed in terms of $E$: 
\be\label{e34}
Q(s)=-\frac{8\pi i}{3\omega \mu_0} h(s) (\nb\times E)(s), \ee 
see \cite{R581}.
 Here the factor $\frac{8\pi}{3}$ appears if $D_m$ are balls.
Otherwise a tensorial factor $c_m$, depending on the shape of $S_m$,
should be used in place of $\frac{8\pi}{3}$. 
The factor $c_m$ is defined 
by the formula $\int_{S_m} \nb \times E_e(s)ds=a^2c_m \nb \times 
E_e(x_m)$,
where $x_m\in S$ is a point in $D_m$.   

Thus, equation \eqref{e34} takes the form 
\be\label{e35} E(x)=E_0(x)-\frac{8\pi i}{3\omega \mu_0}
\nb\times \int_{S}g(x,s)h(s)N(s)\nb\times E(s)ds. \ee

It follows from equation \eqref{e35} that the limiting field $E(x)$ 
satisfies
equation \eqref{e7} away from $S$, and a transmission boundary
condition on $S$:

\be\label{e36}
[N_s, E_-(s)-E_+(s)]=-\frac{8\pi i}{3\omega \mu_0} h(s)N(s)\nb\times
E(s).
\ee

\section{Conclusions}
It is proved that a distribution of many small impedance 
particles in a thin layer on a smooth surface $S$ can be described by a 
transmission boundary condition \eqref{e36}. This condition 
shows that the equivalent surface currents on $S$ are calculated 
analytically in the limit $a\to 0$ in terms of boundary impedance
function $h$ and the distribution density function $N(s)$.
Therefore, these currents can be controlled.

\end{document}